\documentclass[reprint,superscriptaddress,amsmath,amssymb,aps,prl]{revtex4-2}
\usepackage{float}
\usepackage{graphicx} 
\usepackage{dcolumn} 
\usepackage{bm} 
\usepackage{xcolor}

\usepackage{xr-hyper}
\newcommand{\col}[2][red]{%
\begingroup
\color{#1}%
#2%
\endgroup
}

\begin{document}

\title{Optical Manifestations of Quantum Geometry in Electron-Phonon Coupling}

\author{Jiaming Hu}
\affiliation{School of Materials Science and Engineering, Zhejiang University, Hangzhou 310027, China}
\affiliation{Department of Materials Science and Engineering, Westlake University, Hangzhou 310030, China.}
\affiliation{Zhejiang Key Laboratory of 3D Micro/Nano Fabrication and Characterization, Westlake University, Hangzhou 310030, China}

\author{Wenbin Li}
\email{liwenbin@westlake.edu.cn}
\affiliation{Department of Materials Science and Engineering, Westlake University, Hangzhou 310030, China.}
\affiliation{Zhejiang Key Laboratory of 3D Micro/Nano Fabrication and Characterization, Westlake University, Hangzhou 310030, China}

\author{Zhichao Guo}
\affiliation{Center for Quantum Matter, School of Physics, Zhejiang University, Hangzhou 310058, China.}

\author{Hua Wang}
\email{daodaohw@zju.edu.cn}
\affiliation{Center for Quantum Matter, School of Physics, Zhejiang University, Hangzhou 310058, China.}

\author{Kai Chang}
\affiliation{Center for Quantum Matter, School of Physics, Zhejiang University, Hangzhou 310058, China.}

\date{\today}
\begin{abstract}
    Quantum geometry is crucial for understanding intricate condensed matter systems, governing transport phenomena and optical responses. However, traditional studies predominantly consider a static crystal lattice, focusing exclusively on the pure-electronic quantum geometry of the Hilbert space parameterized by electronic wave vectors, thereby overlooking the dynamic effects arising from phonons and their coupling with electrons. In this work, we reveal the intrinsic quantum geometry of the electron-phonon coupling (EPC), which resides in the hybrid Hilbert space parameterized by both the electronic wave vectors and phonon displacements. The EPC quantum metric, EPC Berry curvature and EPC shift vector, as central elements, quantify the EPC-induced velocity, polarization, and anomalous charge-center shift, respectively. We further connect this geometry to phonon-mediated optical responses, particularly in-gap resonances, enabling experimental detection and characterization of EPC quantum geometry. 
\end{abstract}

\maketitle

\textit{Introduction.}---The geometric properties of quantum states are increasingly recognized as critical for understanding emergent phenomena in condensed matter ~\cite{liu2024quantum, ahn2022riemannian,xiao2010berry,nagaosa2010anomalous,moore2010confinement,sundaram1999wave}, including electric transport ~\cite{PhysRevLett.49.405,sodemann2015quantum,gao2014field,citro2023thouless}, and optical responses~\cite{wang2022generalized,PhysRevX.10.041041,gu2023discovery,morimoto2016topological,orenstein2013berry,morimoto2016semiclassical,holder2020consequences,bhalla2022resonant}. Lattice dynamics, which is fundamental to modern solid-state physics, governs a wide range of material properties~\cite{shin2020dynamical,RevModPhys.73.515,giustino2017electron}, including infrared and Raman spectra~\cite{okamura2022photovoltaic,tolles1977review}, electronic polarization~\cite{vanderbilt1993electric,PhysRevB.49.14202}, thermal and electrical transport~\cite{dai2021phonon,giustino2007electron,bardeen1950deformation,zhou2018electron,zhu2022giant}, ferroelectricity~\cite{onoda2004topological,im2022ferroelectricity,shin2021quantum}, piezoelectricity, density waves~\cite{wang2023strong,wang2023strong}, and superconductivity~\cite{giustino2007electron,kulic2000interplay,marsiglio2008electron}. Nevertheless, the interplay between quantum geometry and lattice dynamics remains largely unexplored. Current research on quantum geometry primarily focuses on Hilbert spaces parameterized by either electronic wavevectors~\cite{ahn2022riemannian,xiao2010berry,nagaosa2010anomalous,moore2010confinement,sundaram1999wave} or molecular displacements~\cite{saparov2022lattice,im2022ferroelectricity}, often neglecting the interactions between these two aspects.
A recent study suggests that the quantum geometry of the electron bands plays a significant role in the strength of electron-phonon coupling (EPC)~\cite{yu2024non}, prompting further investigation into the implications of quantum geometry for lattice dynamics.

In this Letter, we first introduce a series of novel EPC quantum geometric quantities through a systematic analysis, including EPC quantum metric, EPC Berry curvature, and EPC shift vector, which quantify the fundamental complex Riemannian geometry in a hybrid Hilbert space parameterized by both electronic wavevectors and phonon displacements. Then we unveil their profound roles in EPC-induced velocity, polarization, and charge-center shift, which are directly connected to phonon-mediated linear and nonlinear optical responses in crystal solids, particularly those exhibiting below-bandgap resonances at phonon frequencies. 

We adopt the following scheme to describe EPC~\cite{giustino2017electron}: Consider the electronic Hamiltonian $\hat{\mathcal{H}}$ in a crystalline solid and the corresponding Bloch Hamiltonian $\hat{H}_{\bm{k}} = e^{i\bm{k}\cdot{\bm{r}}} \hat{\mathcal{H}} e^{-i\bm{k}\cdot{\bm{r}}}$, where $\bm{k}$ represents the electronic wave vector, the cell-periodic part of the electronic eigenstates $|m\bm{k}{\rangle}$ and the corresponding eigenenergies $\epsilon_{m\bm{k}}$ are defined by the equation $\hat{H}_{\bm{k}}|m\bm{k}{\rangle} = \epsilon_{m\bm{k}}|m\bm{k}{\rangle}$. In the following discussions, the explicit $\bm{k}$ index is omitted for notational simplicity. The Hamiltonian $\hat{H}$ depends on the crystal lattice configuration and is expressed as $\hat{H} = \hat{H}(\{\bm{\tau}_{\kappa}\})$, where $\kappa$ indexes the ionic sites. When the lattice undergoes the vibrations of a $\Gamma$-phonon indexed as $s$, which does not break lattice translational symmetry, the ionic positions are displaced as $\bm{\tau}_{\kappa} \rightarrow \bm{\tau}_{\kappa} + Q_s \Delta \bm{\tau}^{s}_{\kappa}$, where $Q_s$ is a scalar with dimensions of length that quantifies the amplitude of the phonon vibration. The eigen-displacement ${\Delta}\bm{\tau}^{s}_{\kappa} = {\rm Re}[\sqrt{M/m_{\kappa}} \bm{e}^{s}_{\kappa}]$ is dimensionless and mass-normalized, with $m_{\kappa}$ denoting the mass of the $\kappa$-th ion, and $M$ serving as a mass constant. The electronic part of the EPC operator is defined as $\hat{g}_s = \bar{Q}_s \partial_s \hat{H}$, where $\bar{Q}_s \equiv \sqrt{\hbar/2M\omega_s}$ is the displacement quantum, $\omega_s$ the phonon frequency and ${\partial_s} \equiv {\partial}/{\partial Q_s}$ the partial derivative with respect to the phonon displacement amplitude. The EPC matrix element between states $m$ and $n$ is then written as $g_{mn}^s = \bar{Q}_s \langle m | \partial_s \hat{H} | n \rangle$.  
\begin{figure}[htb]
    \includegraphics[width=0.48 \textwidth]{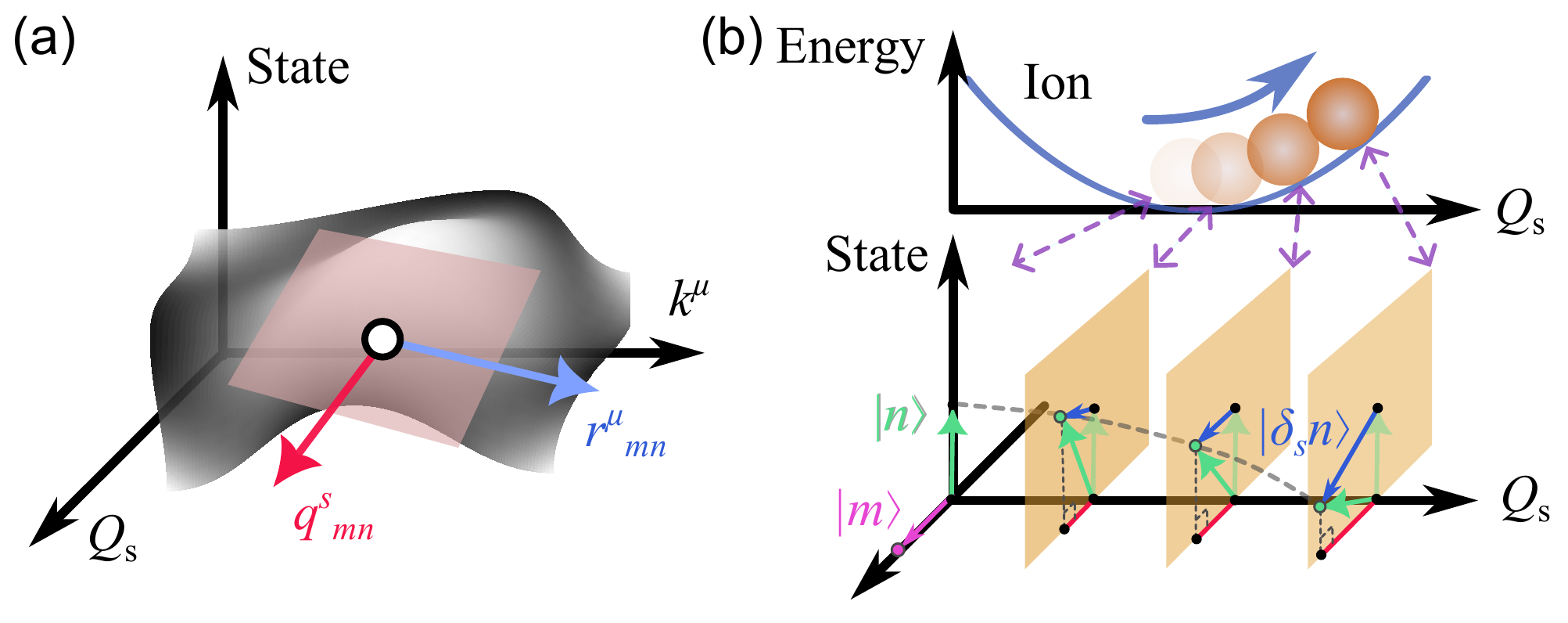}
    \caption{\label{fig:rq_schematic} \textbf{Schematic of EPC tangent vector.} 
    \textbf{(a)} The electronic tangent vector $r^{\mu}_{mn}$ and EPC tangent vector $q^{s}_{mn}$ based on $k^{\mu}$-$Q_s$ parameterization of Hilbert subspace. \textbf{(b)} In phonon-displacement-parametric space. The ionic displacement $Q_s$ (upper panel) varies the crystal configuration, causing the evolution of state vector $|n{\rangle}$ (green arrow, lower panel) by $|{\delta}_sn{\rangle}=|{\partial_s}n{\rangle}Q_s$ (blue arrow), which corresponds to the tangent vector $q^s_{mn}={\langle}m|{\partial_s}n{\rangle}$. } 
\end{figure}

\textit{EPC quantum geometric quantities.---} We begin by introducing the EPC tangent vector, a fundamental building block for the subsequent discussion of EPC geometric quantities. Within the adiabatic approximation, applying $\partial_s$ to the Schrödinger equation $\hat{H}|n\rangle = \epsilon_n|n\rangle$ reveals a fundamental relationship between inter-band EPC matrix element $g_{mn}^s$ and the \textit{EPC tangent vector} $q^s_{mn}$ as (see details in Sec.~III G of Supplemental Material~\cite{SI_info}):
\begin{equation}\label{eq:g_mn_geo}
    iq^s_{mn} = \frac{g^s_{mn}/\bar{Q}_s}{\epsilon_{mn}} = \langle \partial_s m|n\rangle = -\langle m|\partial_s n\rangle,
\end{equation}
while the intra-band term simply becomes $g^s_{mm} = (\partial_s\epsilon_m)\bar{Q}_s$. $q^s_{mn}$ bears close analogy to the optical transition dipole moment with $m{\neq}n$~\cite{parker2019diagrammatic}:
\begin{equation}\label{eq:h_mn_geo}
    ir^{\alpha}_{mn} = \frac{\hbar v_{mn}^{\alpha}}{\epsilon_{mn}} = \langle \partial_{\alpha}m|n\rangle = -\langle m|\partial_{\alpha}n\rangle,
\end{equation}
where $\partial_{\alpha}\equiv\partial/\partial k^{\alpha}$ denotes the partial derivative with respect to the $\alpha$-direction electronic wave vector, and $v_{mn}^{\alpha}$ represents the corresponding velocity matrix element. As schematicized in Fig.~\ref{fig:rq_schematic}, both $r^{\alpha}_{mn}$ and $q^s_{mn}$ act as tangent vectors in the Hilbert subspace spanned by $|m\rangle$ and $|n\rangle$, but differ in their parameterization. 


The complex Riemannian structure of the $k^{\alpha}$-$Q_s$ parametric subspace is naturally defined by the Hilbert-Schmidt inner product of the corresponding tangent vectors~\cite{ahn2022riemannian}:
\begin{equation}\label{eq:G_salpha_mn}
    \mathbb{G}_{mn}^{s;\alpha} = \langle \partial_s m|n\rangle\langle n|\partial_{\alpha}m\rangle = q^s_{mn}r^{\alpha}_{nm},
\end{equation}
which defines the \textit{EPC quantum geometric tensor}. This gauge-invariant quantity represents a novel geometric object distinct from both the pure-electronic $\mathbb{W}^{\alpha\beta}_{mn}=r^{\alpha}_{mn}r^{\beta}_{nm}$~~\cite{ahn2022riemannian,holder2020consequences,resta2011insulating,wang2022generalized,PhysRevX.10.041041} and pure-phononic $\mathbb{M}^{s_1s_2}_{mn}=q^{s_1}_{mn}q^{s_2}_{nm}$ types~\cite{saparov2022lattice,im2022ferroelectricity}. In particular, the imaginary component of $\mathbb{G}_{mn}^{s;\alpha}$, termed the \textit{EPC Berry curvature}, describes the parallel transport of Berry connection $r^{\alpha}_{mm}=i{\langle}m|{\partial_{\alpha}}m{\rangle}$ along $Q_s$ as $-2{\sum_n}\mathrm{Im}[\mathbb{G}_{mn}^{s;\alpha}]=\partial_s [r^{\alpha}_{mm}]$ (modulo a gauge transformation), and has demonstrated general topological significance in ferroelectric polarization~\cite{onoda2004topological}. 

However, the real component of $\mathbb{G}_{mn}^{s;\alpha}$, termed the \textit{EPC quantum metric}, remains largely unexplored. To understand its significance, consider the $k^{\alpha}$-$Q_s$ parameterization as a coordinate map from the Hilbert subspace to $\mathbb{R}^2\ni(k^{\mu}, Q_s)$, where $\mathrm{Re}[\mathbb{G}_{mn}^{s;\alpha}]$ quantifies the `distance' between states $|m\rangle$ and $|n\rangle$. Denoting this distance as $l^{s;\mu}_{mn}$ at parameter point $(k^{\mu}, Q_s=0)$, its variation under an infinitesimal change $(dk^{\mu}, dQ_s)$ follows~\cite{nakahara2018geometry,resta2011insulating}:
\begin{equation}\label{eq:dl2_mn}
    \begin{aligned}
    (dl^{s;\mu}_{mn})^2 &= \mathrm{Re}[\mathbb{W}^{\mu\mu}_{nm}](dk^{\mu})^2 + \mathrm{Re}[\mathbb{M}^{ss}_{nm}](dQ_s)^2 \\
    &\quad + 2\mathrm{Re}[\mathbb{G}^{s;\mu}_{nm}]dk^{\mu}dQ_s,
    \end{aligned}
\end{equation}
which admits two complementary understandings: (i) Symmetry-wise, $\mathrm{Re}[\mathbb{G}^{s;\mu}_{nm}]$ measures the asymmetric response of inter-state distance to phonon displacements: $dl^{s;\mu}_{mn}$ differs for $(dk^{\mu},+dQ_s)$ versus $(dk^{\mu},-dQ_s)$ when $\mathrm{Re}[\mathbb{G}^{s;\mu}_{nm}]\neq 0$, signaling symmetry breaking by phonon displacements. In fact, as shown in Sec.~V E~\cite{SI_info}, this quantity vanishes if with both time-reversal symmetry (TRS) and parity symmetry. (ii) Geometrically, $\mathrm{Re}[\mathbb{G}^{s;\mu}_{nm}]$ is the off-diagonal component of metric tensor, indicating the non-orthogonal relation between $k^{\mu}$ and $Q_s$ in distance measurement, revealing the intertwined dependence of electronic states on $k^{\mu}$ and $Q_s$.  

Based on perspective (ii), this non-orthogonality can be further quantified by introducing the projection angle $\zeta^{s;\mu}_{mn}$. Consider the pure-electronic and pure-phononic variation $dl^{\mu}_{mn}$ and $dl^{s}_{mn}$ induced by parameter shift $(dk^{\mu},0)$ and $(0,dQ_s)$, respectively. The combined variation satisfies $(dl^{s;\mu}_{mn})^2 = (dl^{\mu}_{mn})^2 + (dl^{s}_{mn})^2 + 2(dl^{\mu}_{mn})(dl^{s}_{mn})\cos\zeta^{s;\mu}_{mn}$. Comparing this with Eq.~\ref{eq:dl2_mn} yields: $\zeta^{s;\mu}_{mn} = \arccos\left(\frac{\mathrm{Re}[\mathbb{G}^{s;\mu}_{nm}]}{\sqrt{\mathrm{Re}[\mathbb{W}^{\mu\mu}_{nm}]\mathrm{Re}[\mathbb{M}^{ss}_{nm}]}}\right) = \arg(r^{\mu}_{mn}) - \arg(q^s_{mn}) $, the dispersion of which directly connects to the shift vectors as:
\begin{equation}\label{eq:zeta_R}
    \partial_{\alpha}\zeta^{s;\mu}_{mn} = R^{s;\alpha}_{mn} - R^{\alpha\mu}_{mn},
\end{equation}
where we define the \textit{EPC shift vector} as:
\begin{equation}\label{eq:Rsbeta}
    R^{s;\alpha}_{mn} = - \partial_{\alpha}\arg(q^s_{mn}) + (r^{\alpha}_{mm}-r^{\alpha}_{nn}).
\end{equation}
which is analogous to the well-established electronic shift vector $R^{\alpha\mu}_{mn}=-\partial_{\alpha}\arg(r^{\mu}_{mn})+(r^{\alpha}_{mm}-r^{\alpha}_{nn})$~\cite{wang2022generalized,sipe2000second,sciadv.aav9743,shi2019shift}. Crucially, the EPC shift vector depends solely on the phase and not the EPC strength, referring to the Hermitian connection in EPC-parameteric space~\cite{ahn2022riemannian}. Eq.~\ref{eq:zeta_R} establishes a fundamental relationship between three quantum metrics and two shift vectors, characterizing the $k^{\alpha}$-dependence of $Q_s$-$k^{\mu}$ non-orthogonality. This higher-order EPC geometric effect emerges exclusively in noncentrosymmetric systems (discussed in Sec.~V E~\cite{SI_info}).


\textit{Physical interpretations.}---The EPC quantum geometric quantities introduced above have profound physical implications. The tangent vector $q^s_{nm}$ governs the first-order perturbation of electronic states by EPC as:
\begin{equation}\label{eq:delta_sm}
    |\delta_s m\rangle = \sum_{n\neq m}\frac{\langle n|\hat{g}^s|m\rangle}{\epsilon_{mn}}|n\rangle = 
    -\bar{Q}_s\sum_{n\neq m}iq^s_{nm}|n\rangle,
\end{equation}
where $\epsilon_{mn}=\epsilon_m-\epsilon_n$. 
This perturbation manifests key phenomena that embody EPC quantum geometric tensor. For instance, the EPC quantum metric leads to the EPC-induced velocity:
\begin{equation}\label{eq:Dv_musm}
    \begin{aligned}
        \Delta v^{s;\mu}_m &= \langle \delta_s m|\hat{v}^\mu|m\rangle + \mathrm{h.c.} = 2\sum_{n\neq m}\frac{\epsilon_{mn}}{\hbar}\bar{Q}_s\mathrm{Re}[\mathbb{G}^{s;\mu}_{nm}].
    \end{aligned}
\end{equation}
The EPC Berry curvature, on the other hand, drives EPC-induced polarization:
\begin{equation}\label{eq:DP_sm}
    \Delta P^{s;\alpha}_m = e\langle \delta_s m|\hat{r}^\alpha|m\rangle + \mathrm{h.c.}
    = -2e\bar{Q}_s\sum_n\mathrm{Im}[\mathbb{G}_{mn}^{s;\alpha}].
\end{equation}
Consequently, we can define the dimensionless \textit{EPC character number} as $\mathcal{C}^{s;\mu} = -2\bar{Q}_s\sum_m f_m\int\frac{dk^\mu}{2\pi}\mathrm{Im}[\langle\partial_s m|\partial_\mu m\rangle]$ ($f_m$ is the occupation factor of state $m$), which represents the covalent component of Born effective charge~\cite{resta2023adiabaticobservables,onoda2004topological,fachin2024nearly}, as a specific application of modern polarization theory in EPC~\cite{vanderbilt2018berry}. Further details can be found in Sec.~III H of Supplemental Materials~\cite{SI_info}. 
\begin{table*}[tb]
    \begin{tabular}{lp{3.5cm}p{3.5cm}p{4.5cm}}
    \hline\hline 
        Geometric quantity & Pure-electron & Electron-phonon & Physical significance \\ \hline
        Tangent vector & $r^\alpha_{mn}$ & $q^s_{mn}$ (Eq.~\ref{eq:g_mn_geo}) & Transition probability \\
        Quantum geometric tensor & $\mathbb{W}^{\alpha\beta}_{mn}=r^\alpha_{mn}r^\beta_{nm}$ & $\mathbb{G}^{s;\tau}_{mn}=q^s_{mn}r^\tau_{nm}$ & -- \\
        Quantum metric & $\mathrm{Re}[\mathbb{W}^{\alpha\beta}_{mn}]$ & $\mathrm{Re}[\mathbb{G}^{s;\tau}_{mn}]$ & Optical absorption~\col{\cite{ghosh2024probing}}; EPC-induced velocity \\
        Berry curvature & $\mathrm{Im}[\mathbb{W}^{\alpha\beta}_{mn}]$ & $\mathrm{Im}[\mathbb{G}^{s;\tau}_{mn}]$ & Anomalous velocity~\col{\cite{xiao2010berry}}; EPC-induced polarization \\
        Shift vector & $R^{\alpha\beta}_{mn}$ & $R^{s;\beta}_{mn}$ & Charge center shift (geodesic curvature)\col{~\cite{qian2022role,wang2024geodesicnaturequantizationshift}} \\
        Chern/Character number & $\iint dk^\alpha dk^\beta\mathrm{Im}[\mathbb{W}^{\alpha\beta}]/2\pi$ & $\iint dk^\mu dQ_s\mathrm{Im}[\mathbb{G}^{s;\mu}]/2\pi$ & Anomalous Hall effect~\cite{nagaosa2010anomalous};\col{Born effective charge~\cite{resta2023adiabaticobservables,onoda2004topological,fachin2024nearly}} \\
        Hermitian connection & $r^\alpha_{nm}r^\beta_{mn}R^{\alpha\beta}_{mn}$ & $q^s_{nm}r^\beta_{mn}R^{\alpha\beta}_{mn}$ & Shift current~\cite{tan2016shift,sipe2000second} \\ \hline\hline
    \end{tabular}
    \caption{\label{tab:sum_quantum} Comparison of quantum geometric quantities in pure-electronic versus electron-phonon coupled systems.}
\end{table*}

Beyond intrinsic EPC effects, we consider light-perturbed EPC, modeled via the minimal substitution $\hat{g}^s(\bm{k}) \rightarrow \hat{g}^s(\bm{k} + e\bm{A}(\omega)/\hbar)$, where $\bm{A}(\omega)$ denotes the classical vector potential of light with frequency $\omega$. This perturbation leads to a first-order correction of the form $\delta^\alpha\hat{g}^s = \hat{\mathcal{D}}^\alpha \hat{g}^s \, eA^\alpha(\omega)/\hbar$, where $\hat{\mathcal{D}}^\alpha$ is the covariant derivative~\cite{SI_info,parker2019diagrammatic}. As derived in Sec.~III G of the Supplemental Material~\cite{SI_info}, this results in a characteristic polarization:
\begin{equation}\label{eq:DP_mualpha_sm2}
    \Delta P^{s;\mu\alpha}_{m}\big|_{\mathrm{shift}}(\omega) = 
    \sum_{n\neq m} \frac{eR^{s;\alpha}_{nm}E^\alpha(\omega)}{\hbar\omega}
    \, e\bar{Q}_s \, \mathrm{Im}[\mathbb{G}_{mn}^{s;\mu}],
\end{equation}
where $e\bar{Q}_s \, \mathrm{Im}[\mathbb{G}_{mn}^{s;\mu}]$ serves as the polarization source, and $eR^{s;\alpha}_{nm}E^\alpha(\omega)/\hbar\omega$ acts as the driving force. For comparison, the pure-electronic shift vector $R^{\mu\alpha}_{nm}$ generates a shift current with a similar kernel: $\frac{eR^{\mu\alpha}_{nm}E^{\alpha}(\omega)}{\eta_e} \, e\, \mathrm{Re}[\mathbb{W}^{\alpha\alpha}_{nm}]$ (where $\eta_e$ denotes the electronic linewidth)~\cite{wang2022generalized, cook2017design, young2012first}, capturing the light-induced shift of the electronic charge center. In this context, $R^{s;\alpha}_{nm}$ represents a phonon-induced analogue of the anomalous shift process, thus extending previous studies that focused solely on static crystal lattices~\cite{young2012first, ogawa2017shift, PhysRevX.10.041041}.

The concepts of the EPC quantum geometric tensor and EPC shift vector underscore the pivotal role of quantum geometry in capturing the phase aspects of EPC. Traditional research has predominantly focused on the magnitude of EPC matrix elements, which significantly influence properties such as first-order self-energy and relaxation times, relevant to superconduction~\cite{allen1983theory} and carrier mobility~\cite{shao2013first,marsiglio2008electron,Grimvall1981,giustino2017electron}. However, these studies have largely overlooked the crucial phase information. In contrast, EPC quantum geometric quantities retain this phase information by integrating it with electronic transition dipoles to construct gauge-invariant observables relevant to optical responses. This framework provides deeper insight into the geometric phase-related properties of EPC. A summary of the identified EPC quantum geometric quantities is presented in Table~\ref{tab:sum_quantum}. 

\textit{Optical manifestations.}---Building on the profound physical interpretations of EPC quantum geometric quantities, we explore their implications for phonon-mediated linear and nonlinear optical (Ph-LO and Ph-NLO) responses. These responses are derived rigorously using a Feynman diagrammatic approach, as detailed from Sec.~I to~III F in the Supplementary Material~\cite{SI_info}. Physically, the interaction of the EPC-induced polarization ${\Delta}P^{s;\alpha}= \int [d\bm{k}] \sum_m f_m {\Delta}P^{s;\alpha}_{m}$ with optic electric field $E^{\alpha}(\omega)$ facilitates energy transfer from light into phonons, thereby exciting their propagation. Resonant behavior emerges through the `shift mechanism' when the light frequency $\omega$ matches the phonon eigenfrequency $\omega_s$, characterized by the imaginary part of the phonon propagator $-i\pi{\delta}(|\omega|-\omega_s)$ (see Sec.II A and~III C~\cite{SI_info}). The excited phonons can subsequently polarize electrons through either intrinsic EPC or light-perturbed EPC. The former leads to a light-independent polarization ${\Delta}P^{s;\mu}$, and gives rise to the Ph-LO susceptibility as (Sec.~III G~\cite{SI_info}):
\begin{equation}\label{eq:chi_alpha}
    \chi_{\mu\alpha}(\omega) = 
    \frac{i\pi}{\epsilon_0}
    \sum_s
    {\Delta}P^{s;\mu}
    {\Delta}P^{s;\alpha}
    \delta(|\omega| - \omega_s), 
\end{equation}
which can be classically understood through a charged harmonic oscillator discussed in Sec.~V M~\cite{SI_info}. In contrast, the latter leads to a light-dependent polarization $\Delta P^{s;\mu\beta}(\omega_2) = \int [d\bm{k}] \sum_m f_m {\Delta}P^{s;\mu\beta}_{m}(\omega_2)\big|_{\mathrm{shift}}$, corresponding to the second-order Ph-NLO susceptibility (Sec.~III G~\cite{SI_info}):
\begin{equation}\label{eq:chi_alphabeta}
    \begin{aligned}
        &\chi_{\mu\alpha\beta}(\omega;\omega_1,\omega_2) E^{\beta}(\omega_2) = 
        \\
        &-\frac{i\pi}{\epsilon_0}
        \sum_s
        {\Delta}P^{s;\mu\beta}(\omega_2)
        {\Delta}P^{s;\alpha}
        \delta(|\omega_1| - \omega_s).
    \end{aligned}
\end{equation}
As a representative example, consider the phonon-mediated electro-optic effect, where one of the input fields acts as a quasi-static electric field with frequency $\delta\omega \ll \omega$. To highlight the geometric origin, we focus on the linearly polarized electro-optic susceptibility (Sec.~III F~\cite{SI_info}):
\begin{equation}
    \begin{aligned}
        \chi_{\alpha\alpha\alpha}(\omega;\omega,\delta\omega) &=
        \frac{-i\pi e^2}{2\epsilon_0 \omega (\hbar \delta\omega)}
        \sum_s
        \delta(\omega - \omega_s)\Delta P^{s;\alpha}
        \\
        &{\times}
        \left(\Gamma^{\mathrm{curl}}_s + \Gamma^{\mathrm{metric}}_s\right),
    \end{aligned}
\end{equation}
which contains contributions from both the EPC Berry curvature and quantum metric:
\begin{equation}
    \begin{aligned}
        \Gamma^{\mathrm{curl}}_s &= 
        \sum_{mn} \int [d\bm{k}]
        f_{mn}
        \bar{Q}_s \, \mathrm{Im}[\mathbb{G}^{s;\alpha}_{mn}]
        \frac{\epsilon_{mn}}{\hbar} R^{\alpha\alpha}_{mn},
        \\
        \Gamma^{\mathrm{metric}}_s &= 
        \sum_{mn} \int [d\bm{k}]
        f_{mn}
        2\frac{\bar{Q}_s \epsilon_{nm}}{\hbar}
        \mathrm{Re}[\mathbb{G}^{s;\alpha}_{mn}]
        \partial_{\alpha} \ln |h^{\alpha}_{mn}|.
    \end{aligned}
\end{equation}
Thus, $\chi_{\alpha\alpha\alpha}(\omega;\omega,\delta\omega)$ serves as a promising probe of fundamental EPC geometric properties. Furthermore, phonon-mediated shift current and second-harmonic generation, as two other representative second-order Ph-NLO effects, are derived and discussed in Sec.~III D and Sec.~III E~\cite{SI_info}, further illustrate the rich diversity of phonon-mediated optical phenomena and the optical manifestations of EPC quantum geometry.


\textit{Discussions.}---It is noteworthy that, as discussed in Sec.~II B and~II E~\cite{SI_info}, phonon mediation involves two distinct mechanisms: shift~\cite{okamura2022photovoltaic,gu2023discovery} and ballistic~\cite{noffsinger2012phonon,monserrat2018phonon}. A key advantage of shift mechanism lies in its pure-phononic resonance (e.g., $\delta(\omega-\omega_s)$ in Eqs.~\ref{eq:chi_alpha} and~\ref{eq:chi_alphabeta}), which can be activated by photon at the phonon energy scale, often lying below the electronic band gap~\cite{okamura2022photovoltaic}. As discussed in Sec.~V C~\cite{SI_info}, its strength is possibly comparable to pure-electronic responses. This allows shift-type Ph-NLO responses to be clearly distinguished, offering a direct pathway to probe EPC quantum geometry. Furthermore, this mechanism holds particular promise for terahertz optics, providing an additional response channel on the infrared side of the spectrum. For these reasons, we focus on the shift mechanism in the main text, while leaving the discussion of the ballistic mechanism to Secs.~II A and~III I of the Supplemental Material~\cite{SI_info}. Our results reveal an analogy between electron-light and electron-phonon interactions: phonon-induced crystal fields mimic optical electric fields, enabling EPC induced optical effects to be captured through geometric quantities such as the EPC Berry curvature and quantum metric. This connection can be further exemplified by the similarity and discrepancy between pure-electronic and Ph-LO susceptibility (detailed in Sec.~III H~\cite{SI_info}). Consequently, Ph-NLO responses emerge as powerful and sensitive probes of the quantum geometric properties of EPC.

\begin{equation}
    h^{\alpha}_{mn} = \langle m| \partial_{\alpha} \hat{H} | n \rangle 
\end{equation}

\begin{figure}[tb]
    \includegraphics[width=0.46 \textwidth]{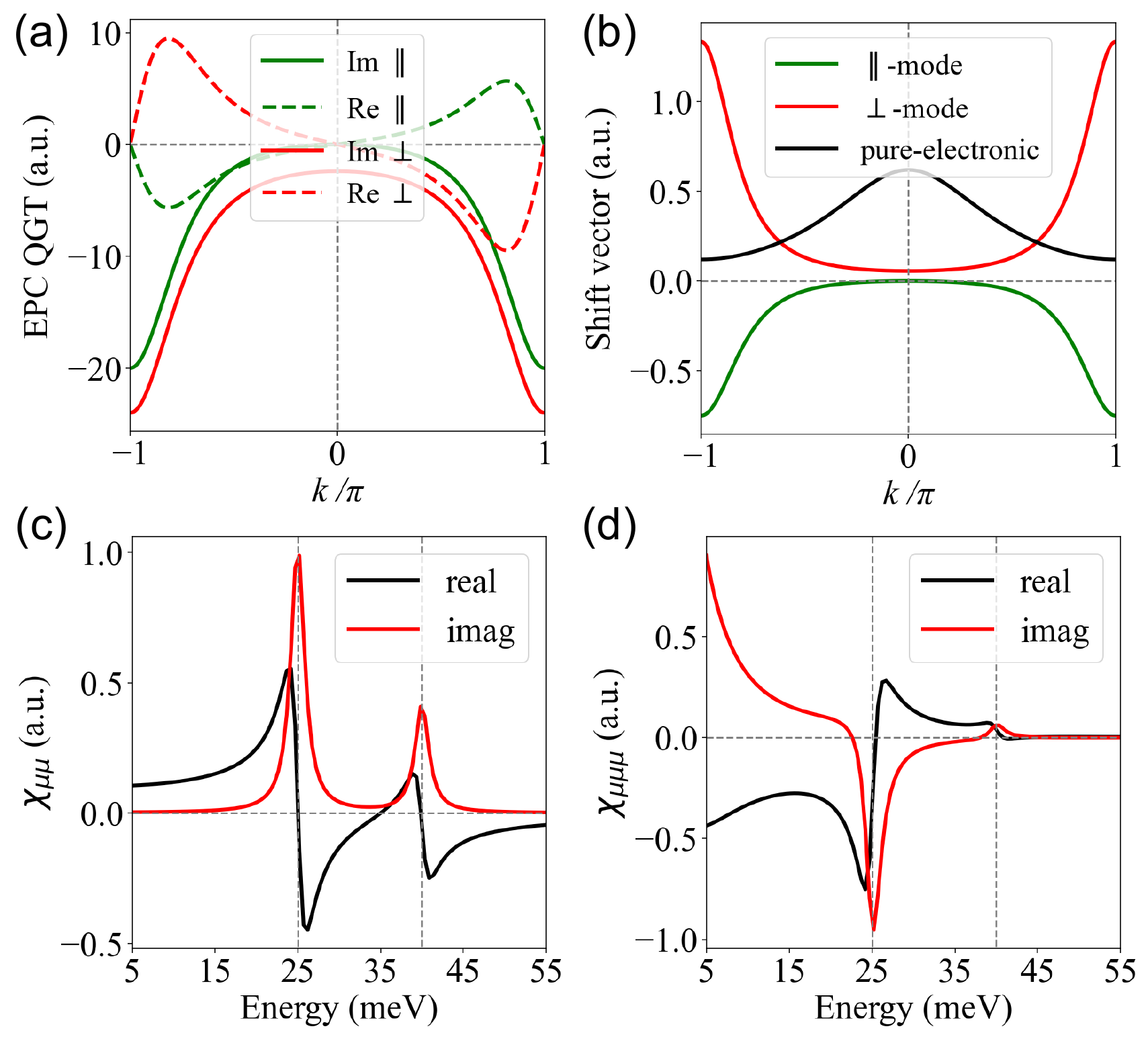}
    \caption{\label{fig:model_results} \textbf{Results of general Rice-Mele model.} \textbf{(a)} EPC quantum geometric tensor (QGT), with real and imaginary components as EPC quantum metric and Berry curvature respectively. \textbf{(b)} Shift vectors of pure-electronic, in-plane-mode ($\parallel$-mode) EPC, and out-of-plane-mode ($\perp$-mode) EPC types.  \textbf{(c)} Phonon-mediated linear susceptibility $\chi_{\mu\mu}(\omega;\omega)$. \textbf{(d)} Phonon-mediated electro-optic susceptibility (Ph-EO) $\chi_{\mu\mu\mu}(\omega;\omega,10^{-3}\omega)$. The electronic energy gap is around 1.0~eV. The $\perp$-mode and $\parallel$-mode phonon energies are 25~meV and 40~meV, respectively. Other used parameters are given in Sec.~IV~\cite{SI_info}. }
\end{figure}

\textit{Model example.}---We further apply the general theory to a dynamic Rice-Mele model, which serves as an effective two-site, one-dimensional description for a wide range of realistic materials (Sec.~IV A~\cite{SI_info}). The staggered potential $\delta_m$ accounts for the influence of substrate, external field or ion species; the static distortion $\gamma$ modifies the inter-site hopping from $u$ to $u\pm \gamma u$, which can represent ferroelectric displacement, double bonding or charge density wave. Additionally, we incorporate the EPC with two optical phonon modes at $\Gamma$ point: the in-plane mode modifies the inter-site distance (denoted as $\parallel$ with phonon displacement $Q_{\parallel}$), creating a dynamic distortion $\pm vQ_{\parallel}$; the out-of-plane mode (denoted as $\perp$ with phonon displacement $Q_{\perp}$) alters the distance between two inequivalent sites and the substrate, hence modifying $\delta_m$ by $wQ_{\perp}$. As a result, the total Hamiltonian becomes: 
\begin{equation}\label{eq:RM_Hamiltonian}
    \hat{H}_k(Q_{\parallel},Q_{\perp}) = 
    \begin{bmatrix}
        \delta_m+wQ_{\perp} & V_{k} \\
        V_{k}^* & -(\delta_m+wQ_{\perp})
    \end{bmatrix},
\end{equation}
with $V_k=u\cos{\frac{ka}{2}}-i(\gamma u + vQ_{\parallel})\sin{\frac{ka}{2}}$. Using representative parameters (Sec.~IV C~\cite{SI_info}), we numerically solve for the EPC quantum geometric tensor and shift vectors. As shown in Fig.~\ref{fig:model_results}(a,b), the EPC shift vector shows comparable strength with pure-electronic type; EPC quantum metric and Berry curvature exhibit odd and even parity with respect to $k$, respectively (Sec.~V E~\cite{SI_info}). The Ph-LO susceptibility $\chi_{\mu\mu}(\omega)$ and Ph-NLO electro-optic susceptibility $\chi_{\mu\mu\mu}(\omega;\omega,10^{-3}\omega)$ are shown in Fig.~\ref{fig:model_results}(c) and (d) respectively, where phonon resonances appear as sub-gap peaks at photon energies of 25 and 40~meV below the electronic energy gap of 1.0~eV. 


\textit{Conclusions.}---By unveiling the EPC tangent vector within the EPC matrix element, we introduce the EPC quantum geometric tensor and the EPC shift vector as essential quantum geometric quantities for both intrinsic and optically perturbed EPC. They transcend traditional frameworks by explicitly incorporating dynamic lattice effects and capturing the phase information of EPC. Their physical interpretation as phonon-induced velocity, polarization, and charge-center shift highlights the fundamental interplay between the electronic system and lattice dynamics. This geometric perspective is anticipated to exert a significant influence on our understanding of a wide range of physical phenomena involving EPC, including ferroelectric phase transitions~\cite{hu2025quantumgeo}, thermal transport, and superconductivity. 

As a representative application, we establish a direct link between these EPC geometric quantities and experimentally observable phonon-mediated optical responses, systematically derived through Feynman diagrammatic method~\cite{SI_info}. These findings pave the way for advancements in terahertz optics and offer a framework for exploring EPC quantum geometry through pure-phononic resonances, complementing existing quantum geometry measurement techniques~\cite{kang2024measurements,PhysRevLett.133.036204}. The theoretical insights, demonstrated with a dynamic Rice-Mele model, are readily adaptable to first-principles calculations and exhibit broad applicability across diverse systems.

\textit{Acknowledgment.}---The authors thank Dr. Zhuocheng Lu from the School of Physics, Zhejiang University for insightful discussions. The work of J.H. and W.L. is supported by the National Natural Science Foundation of China (NSFC) under Project No. 62374136. H.W. acknowledges the support from the NSFC under Grants Nos. 12304049 and 12474240, as well as the support provided by the Zhejiang Provincial Natural Science Foundation of China under grant number LDT23F04014F01. K. C. acknowledges the support from the Strategic Priority Research Program of the Chinese Academy of Sciences (Grants Nos. XDB28000000 and XDB0460000), the NSFC under Grants Nos. 92265203 and 12488101, and the Innovation Program for Quantum Science and Technology under Grant No. 2024ZD0300104.

\bibliography{references}
\end{document}